\documentclass[reprint,showkeys]{revtex4-1}
\usepackage{graphicx}
\usepackage{amssymb}
\usepackage{amsmath}
\usepackage{epstopdf}
\newcommand{\be}{\begin{equation}}\newcommand{\ee}{\end{equation}}
\newcommand{\ba}{\begin{array}{l}}\newcommand{\ea}{\end{array}}
\newcommand{\baa}{\begin{eqnarray}}\newcommand{\eaa}{\end{eqnarray}}
\newcommand{\re}[1]{(\ref{#1})}
\newcommand{\ci}[1]{\cite{#1}}

\begin{document}

\title{Classical and quantum dynamics of a kicked relativistic particle in a box}

\author{J.R. Yusupov, D.M. Otajanov, V.E. Eshniyazov and D.U. Matrasulov}

\affiliation{Turin Polytechnic University in Tashkent,\\ 17 Niyazov Str.,
100095, Tashkent, Uzbekistan}

\begin{abstract}
We study classical and quantum dynamics of a kicked relativistic
particle confined in a one dimensional box. It is found that in
classical case for chaotic motion the average kinetic energy grows
in time, while for mixed regime the growth is suppressed. However,
in case of regular motion energy fluctuates around certain value.
Quantum dynamics is treated by solving the time-dependent Dirac
equation with delta-kicking potential, whose exact solution is
obtained for single kicking period. In quantum case,  depending on
the values of the kicking parameters, the average kinetic energy
can be quasi periodic, or fluctuating around some value. Particle
transport is studied by considering spatio-temporal evolution of
the Gaussian wave packet and by analyzing the trembling motion.
\end{abstract}

\maketitle

\section{Introduction}

Particle dynamics in confined quantum systems has attracted much
attention  in the context of  nanoscale physics
\cite{nanoph1}-\ci{Richter} and quantum chaos theory
\cite{Nak}-\cite{Gutz}. Usually, studies of confined quantum
dynamics within the chaos theory have been focused on two types of
problems. First type deals with the analysis of the spectral
statistics (so-called quantum chaos) by solving Schrodinger
equation in confined geometries (quantum billiards)
\cite{Gutz}-\cite{Uzy}. Second type deals with the  quantum
dynamics in periodically driven systems by studying average
kinetic energy as a function of time \cite{Casati}-\cite{Izr}.

Despite the fact that confined quantum systems are widely studied
in the literature, most of the researches are mainly focused on
the nonrelativistic systems. In this paper we address the problem
of delta-kicked  relativistic particle confined in a
one-dimensional box. Nonrelativistic counterpart of such system
have been considered earlier in classical and quantum chaos
contexts by considering kicked particle in infinite square well
\cite{Well,Roy,Kilbane}. For kicked systems, the main feature of
the dynamics is the diffusive growth of the average kinetic energy
as a function of time in classical case and its suppression for
corresponding quantum system. The latter is called quantum
localization of classical chaos \ci{Casati,Izr}. The dynamics of
the kicked nonrelativistic system is governed by single parameter,
product of the kicking strength and kicking period.  However, as
we will see in the following, the dynamics of the relativistic
system is completely different than that of its nonrelativistic
counterpart: There is no single parameter which governs the
dynamics.

Usually, confined relativistic quantum systems appear in particle
physic models such as MIT bag model \cite{MIT} and the quark
potential models \cite{Mukherjee}. However, recent progress made
in fabrication of graphene and studying its unusual properties
made possible experimental realization of Dirac particle confined
in one- \ci{Brey,Matrasulov} and two-dimensional boxes
\cite{Beenakker,Neto,Cortijo}. Such condensed matter realization
of a Dirac particle in a box can be also realized in graphene
nanoribbon ring \cite{GR1,GR2,GR3,GR4,GR5,GR6,GR7} or dot
\cite{Basak,Ramos} which is extensively studied recently both
theoretically and experimentally. Graphene nanoribbon is a strip
of graphene having different edge geometries. The quasiparticle
dynamics in such material is effectively one-dimensional, i.e. can
be described by one dimensional Dirac equation \ci{Brey}. When its
length is finite, it becomes "Dirac particle in a 1D box".
 ``Kicked'' version of such system, i.e., kicked Dirac
particle in a box can be realized, e.g.,  by putting it in a
standing laser wave. One of such models has been recently studied
in \ci{Matrasulov} by focusing on transport phenomena.

We note that earlier, the Dirac equation for a particle confined
in a box was considered in detail in the Refs.
\cite{Alonso}-\cite{Menon}. Unlike the Schr\"odinger equation for
a box, introducing confinement in the Dirac equation via infinite
square well or box faces some difficulties caused by the Klein
tunneling and the electron-positron pair creation \cite{Greiner}.
To avoid such complication, in the Ref. \cite{Berry} the authors
considered the situation when confinement is provided by a
Lorentz-scalar potential, i.e. by a potential coming in the mass
term. Such a choice of confinement is often used in MIT bag model
\cite{MIT} and the potential models of hadrons \cite{Mukherjee}.
Another way to avoid this complication is to impose box boundary
conditions in such a way that they provide zero-current and
probability density at the box walls.
 In the Ref. \cite{Alonso} the types of
the box boundary conditions, providing vanishing current at the
box walls and keeping the Dirac Hamiltonian as self-adjoint are
discussed.

The paper is organized as follows. In the next section we consider
classical dynamics of a relativistic particle confined in a one
dimensional box. In section 3, following the Ref. \cite{Alonso},
we briefly recall the problem of stationary Dirac equation for one
dimensional box. In section 4 we treat the time-dependent Dirac
equation with delta-kicking potential with the box boundary
conditions. In section 5 we discuss wave packet dynamics and
trembling motion. Finally, section 6 presents some concluding
remarks.

\section{Classical dynamics}

Classical relativistic particles whose motion is spatially
confined may appear in plasma \cite{Tajima} and astrophysical
systems \cite{Rogava}. Confinement in such systems can be provided
by constant electric or magnetic fields. Hamiltonian of a delta
kicked relativistic particle is given by (in the units $m=c=1$)
\begin{equation}
H=\sqrt{p^2 +1} -\varepsilon\cos\left(\frac{2\pi}{\lambda} x
\right)\sum_l\delta(t-lT),
\end{equation}
where $\lambda$ is the wavelength, $\varepsilon$ and $T$ are the kicking
strength and period, respectively.

Classical dynamics of a relativistic particle in one-dimensional space is
governed by Hamiltonian equations which are given as
\begin{equation*}
\frac{dp}{dt} =\frac{\partial H}{\partial x},
\end{equation*}

\begin{equation}\label{class}
\frac{dx}{dt} =-\frac{\partial H}{\partial p}.
\end{equation}
Assuming that the motion of the particle is confined within the
box of size $L$ and solving Eqs.(\ref{class}) by imposing
box-boundary conditions, one can analyze the classical dynamics of
a relativistic particle confined in a 1D box. Nonrelativistic
counterpart of this problem was studied in the Refs.
\cite{Well,Kilbane}, where the map describing phase-space
evolution of the system is derived. Unlike the kicked rotor,
classical dynamics of a kicked particle in a box depends not only
on the product of the kicking strength and period, but also on the
number of pulse waves in the box, i.e. on the ratio of the
wavelength to the box size \cite{Well,Kilbane}. In other words,
kicked particle confined in a box has much larger parametric space
than that for kicked rotor. Relativistic generalization of the
map  derived in \cite{Well} can be written as
%\begin{eqnarray}
%\nonumber x_{n+1}&=& \left[ L + (-1)^{B_n}\left( x_n +
%\frac{Tp_n}{\sqrt{p_n^2+1}}-
%\mbox{Sgn}{(p_n)}LB_n \right) \right]\ mod \ L, \\
%\label{map} p_{n+1} &=& (-1)^{B_n}p_n + \frac{2\pi\varepsilon
%T}{\lambda}\sin{\left(\frac{2\pi}{\lambda} x_{n+1} \right)},
%\end{eqnarray}
\begin{eqnarray}
\nonumber x_{n+1} &=& \left.\left( L + (-1)^{B_n}\right( x_n + Tp_n\right/\\
\nonumber \ & & \left. \left.\sqrt{p_n^2+1} - \mbox{Sgn}{(p_n)}LB_n A \right) \right)\ mod \ L, \\
\label{map} p_{n+1} &=& (-1)^{B_n}p_n + \frac{2\pi\varepsilon
T}{\lambda}\sin{\left(\frac{2\pi}{\lambda} x_{n+1} \right)},
\end{eqnarray}
where $B_n=\left[ \mbox{Sgn}(p_n)(x_n + p_n)/L \right]$; $[...]$
is the number of bounces of the particle between the walls during
the interval between $n$th and $(n +1)$th kick and Sgn(...) stand
for integer part and sign of the argument respectively.  It is
clear that this map (as the equations of motion themselves) is
Lorentz invariant, since it is a discretized version of Eqs.
\re{class}.

\begin{figure}[htb]
\includegraphics[width=86mm]{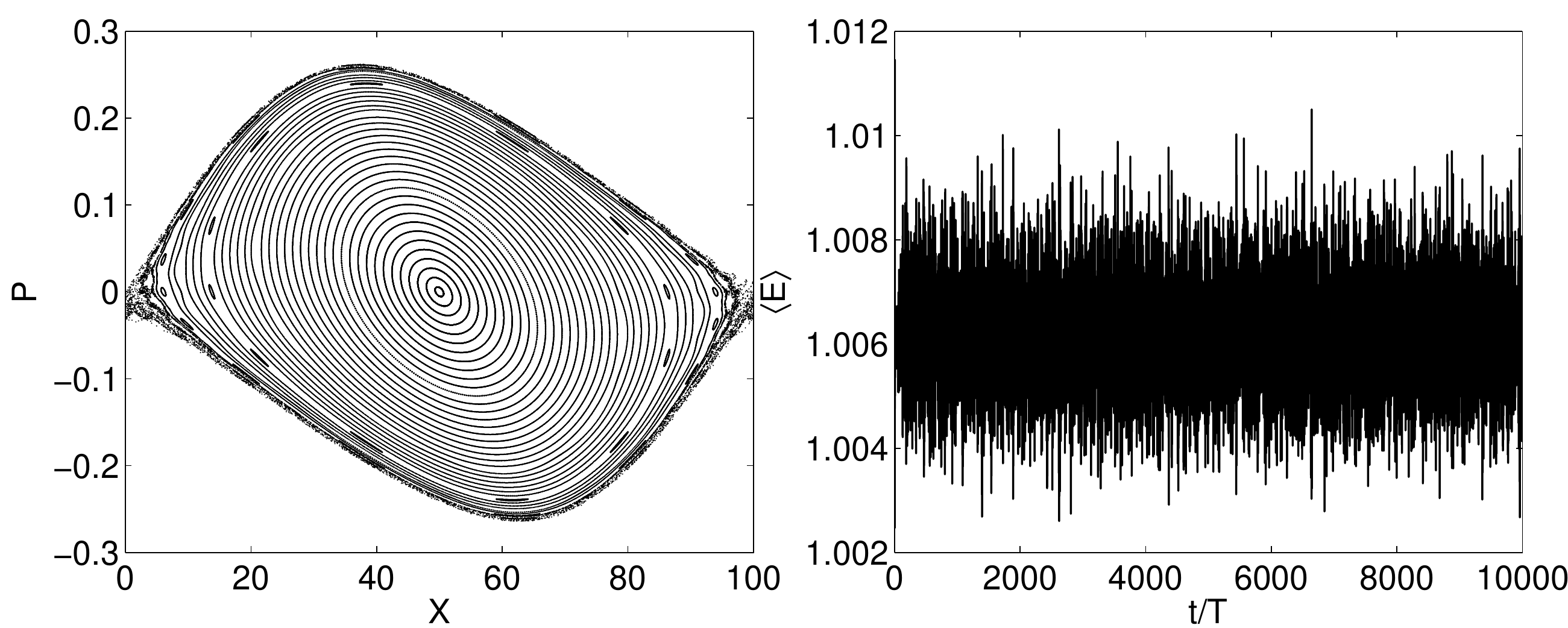}
\caption{Phase-space portrait (a) and the time-dependence of the
average energy vs the number of kicks (b) for the classical
system. The kicking strength $\varepsilon=0.0159$ and the kicking
period $T=100$.} \label{phsp1}
\end{figure}

\begin{figure}[htb]
\includegraphics[width=86mm]{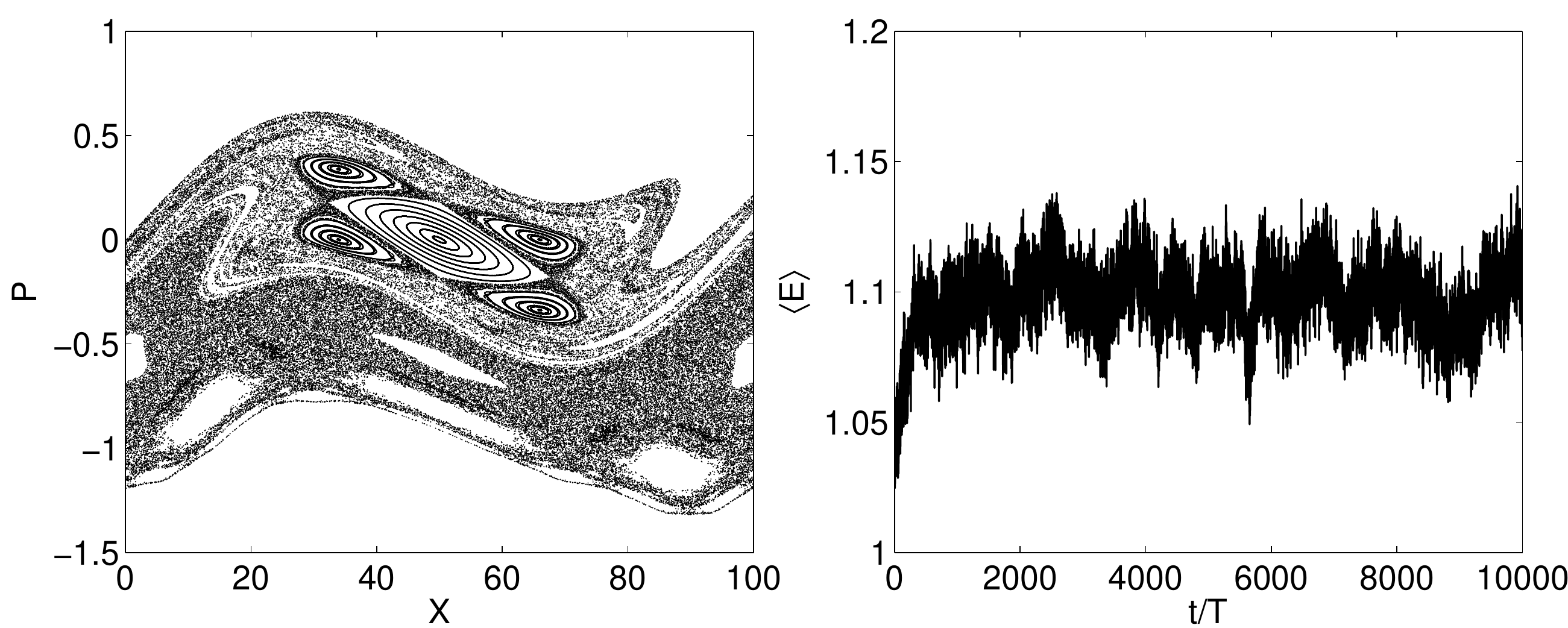}
\caption{The same as in Fig.\ref{phsp1} for $\varepsilon=0.0637$ and $T=100$.}
\label{phsp2}
\end{figure}

\begin{figure}[htb]
\includegraphics[width=86mm]{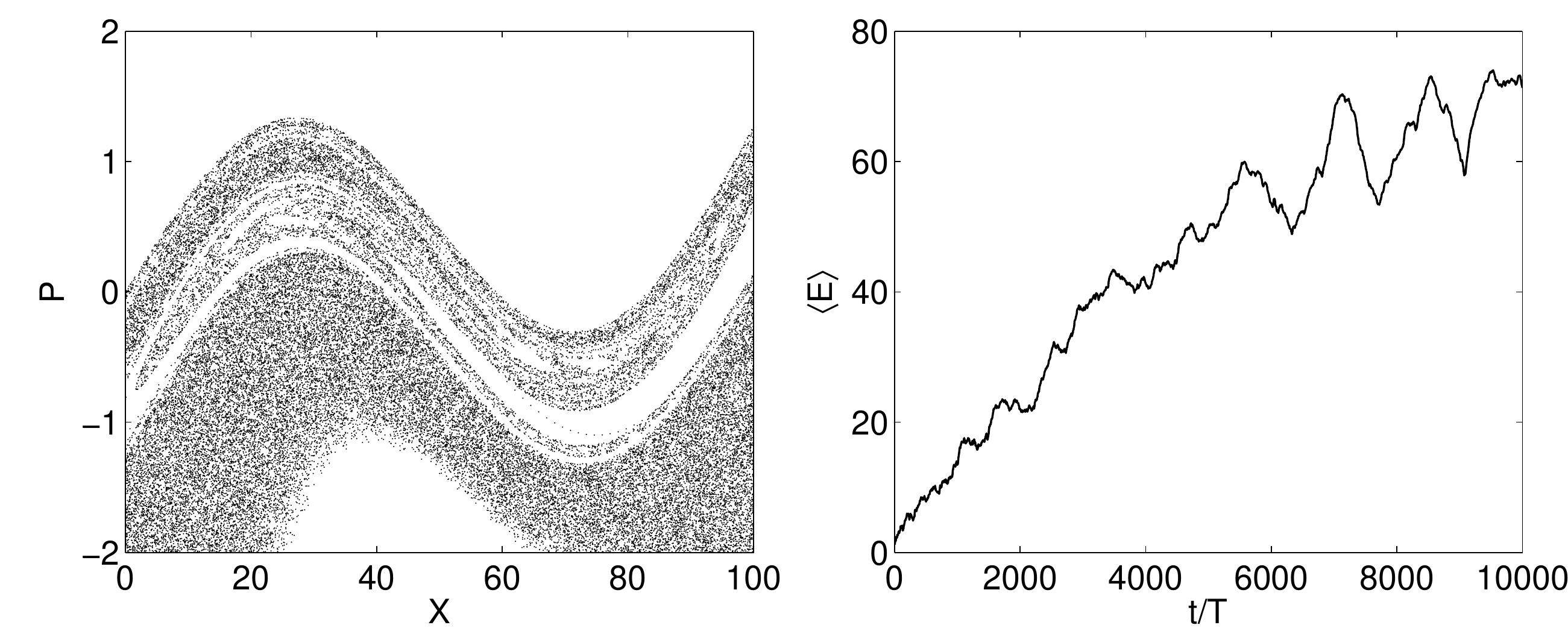}
\caption{The same as in Fig.\ref{phsp1} for $\varepsilon=0.1751$ and
$T=99.99327$.} \label{phsp3}
\end{figure}

\begin{figure}[htb]
\includegraphics[width=75mm]{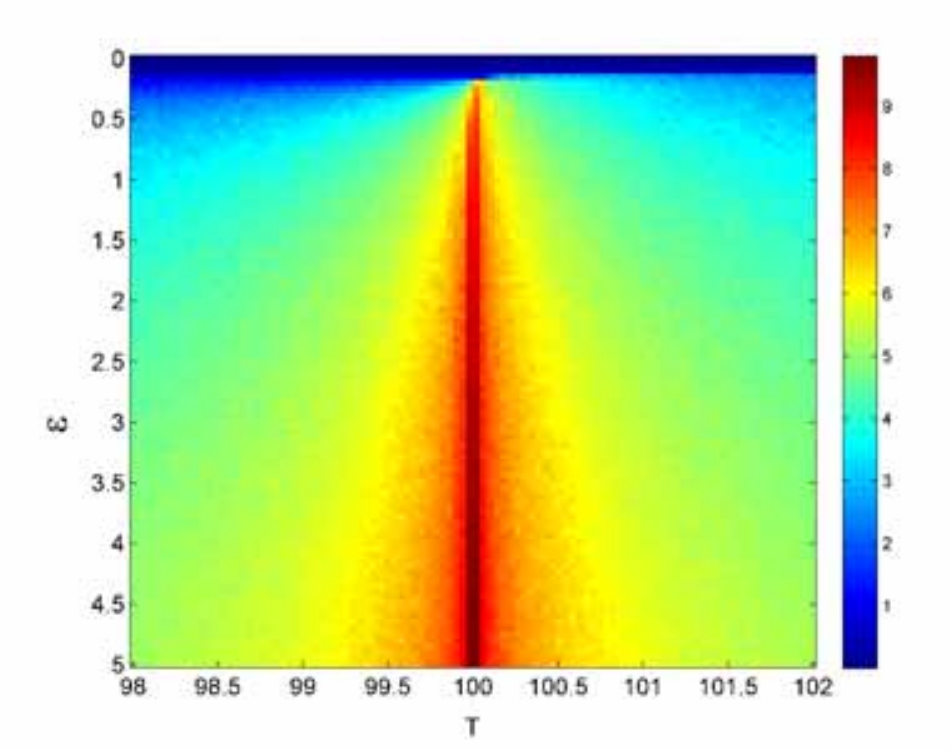}
\caption{(Color online) Average kinetic energy as a function of the kicking
parameters at the 1000th kick (logarithmic scale).} \label{phsp4}
\end{figure}

Fig.\ref{phsp1} presents phase-space portraits (a) and the average
kinetic energy (b) of a kicked relativistic particle in a 1D box
for those values of the kicking parameters at which the dynamics
of the particle is regular. For this case the average kinetic
energy does not grow in time (unlike the nonrelativistic case) and
fluctuates around some fixed value. In Fig.\ref{phsp2} similar
plot is presented for the values of the kicking parameters causing
mixed dynamics. The energy in Fig. \ref{phsp2} (b)grows in time
and the growth is suppressed after the certain number of kicks. In
Fig.\ref{phsp3} phase-space portrait (a) and time-dependence of
the average kinetic energy (b) are plotted for fully chaotic case.
The energy grows almost monotonically for this case in the
considered time period. This regime can be considered as an
acceleration mode. Existence of the acceleration modes can be
clearly seen from the Fig.\ref{phsp4}, where the average kinetic
energy is plotted vs kicking parameters, $\varepsilon$ and $T$.
Maximum values of $E(t)$ and acceleration are possible around
certain values of the kicking period $T=100$ while the growth of
$E(t)$ is not ``sensitive'' with respect to the values of
$\varepsilon$. Our numerical experiments showed that the growth of
$E(t)$ depends on the parameters $K=2\pi\varepsilon T/\lambda$.

We note that there is no single parameter governing the dynamics
of the system for relativistic kicked particle in a box. In other
words,  scaling of the map \re{class} in such way that it will
depend on single parameter, $K=\varepsilon T$, is not possible.
Instead, it depends on each parameter separately which makes the
dynamics of the system more rich than that of the nonrelativistic
counterpart. As shows  the Fig.\ref{phsp4}, the average energy and
dynamics are more sensitive to the change of the kicking period,
$T$ than the kicking parameter,$\varepsilon$. Such a feature can
be related to the fact that the underlying factor for the growth
of the average kinetic energy is the "correlations" between the
bouncing of the particle on the  box walls and the kicks. Indeed,
the dynamics of the particle depends on two interactions, bouncing
from the walls and on the kicking pulses. In case, if the kicks
and bounces are "synchronized", the dynamics can be regular.
Otherwise, chaotization of the system may occur. Moreover, our
detailed numerical analysis of the map and phase-space portraits
showed that unlike the classical  nonrelativistic kicked particle
in a box studied in \cite{Well,Kilbane}, for
(classical)relativistic kicked particle in a box, the diffusion
and growth of the energy does not occur for all values of the
kicking parameters. Namely, diffusive growth is possible only for
fully chaotic regime, while for regular dynamics the average
kinetic energy does not grow at all.

\section{Dirac particle in a one dimensional box}

Unperturbed version of the system we are going to treat, i.e.,
Dirac particle confined in a one-dimensional box was considered
earlier in the Refs. \cite{Alonso}-\cite{Roy}. Here we briefly
recall the description of such system following the Ref.
\cite{Alonso}. In the nonrelativistic  case the box boundary
conditions for the Schr\"odinger equation are introduced through
the infinite square well, either  by imposing the boundary
conditions providing zero-current at the box walls. In case of the
Dirac equation introducing of infinite well leads the Klein
paradox and to electron-positron  pair production from vacuum
\cite{Berry,Greiner}. The latter implies that the problem cannot
be treated within the one-particle Dirac equation that makes
impossible using the infinite well based description of the
particle-in-box system \cite{Berry,Greiner}. Instead, the box can
be introduced through the boundary conditions, providing
zero-current at the boundary and keeping the self-adjointness of
the problem. In case of the Schr\"odinger equation the boundary
condition, $\psi=0$ keeps the Schr\"odinger operator as
self-adjoint. However, for Dirac equation the boundary conditions
at the box walls should be determined carefully
\cite{Alonso}-\cite{Alonso2} to keep the self-adjointness of the
problem.

The stationary Dirac equation for free particle  in  a one-dimensional box
given on the interval $(0, L)$ can be written as (in the system of units $m
=\hbar =c=1$)
\begin{equation}
H_0\psi=\left(-i\alpha_x \cdot \frac{d}{dx} + \beta
\right)\psi=E\psi, \label{dirac_eq}
\end{equation}
where $\alpha_x$ and $\beta$ are the  Dirac matrices. The wave function, $\psi$
can be written in two component form as
\begin{equation}
\psi=\left( \begin{array}{c}  \phi \\ \chi \end {array} \right),
\end{equation}
where large, $\phi$ and small, $\chi$ components are  also two-component
semi-spinors:
$$
\phi=\left( \begin{array}{c}  \phi_1 \\ \phi_2 \end {array}
\right)\;\;\;\chi=\left( \begin{array}{c}  \chi_1 \\ \chi_2 \end
{array} \right),
$$
respectively.

The system of first order differential equations (\ref{dirac_eq}) can be
reduced to second order, Helmholtz-type equation by eliminating one of the
components:
\begin{equation}
\left( \frac{d^2}{dx^2}+k^2 \right) \phi_i=0\;\;\; i=1,2
,\label{large_comp1}
\end{equation}
Here
\begin{equation}
k=(E^2- 1)^{1/2}. \label{eigenvalue}
\end{equation}
The small and large components are related to each other through the expression
\begin{equation}
\left( \begin{array}{c}  \chi_1 \\ \chi_2 \end {array}
\right)=\frac{-i}{E+1}\left( \begin{array}{cc} 0 & \frac{d}{dx} \\
\frac{d}{dx}  & 0  \end {array} \right) \left( \begin{array}{c} \phi_1 \\
\phi_2 \end {array} \right). \label{comps_rel}
\end{equation}
Therefore box boundary conditions should be imposed for one of the
components only. Then, one of the positive energy solutions can be
obtained by taking $\phi_2=0$ and therefore $\chi_1=0$. Thus the
general solution for $\phi_1$ can be written as
\begin{equation}
\phi_1=A_1 e^{ikx}+B_1e^{-ikx},
\end{equation}
where $A_1$ and $B_1$ are complex constants. For $\chi_2$ one can obtain
\begin{equation}
\chi_2=\frac{k}{E+ 1}\left( A_1 e^{ikx}-B_1e^{-ikx} \right).
\end{equation}

Imposing the box boundary conditions given by \cite{Alonso},
\begin{equation}
\phi_1(0)=\phi_1(L)=0, \label{condition1}
\end{equation}
we get the following eigenfunctions \cite{Alonso}:
\begin{equation}
\psi_n=2A_n\left( \begin{array}{c}  i \sin(k_nx) \\ 0 \\ 0 \\
\frac{k_n}{E_n+ 1} \cos(k_nx) \end {array} \right), \label{ps1}
\end{equation}
where $E_n$ determined by Eq.(\ref{eigenvalue}), $A_n$ is the
normalization constant,  and $k_n=\pi n/L,n=1,2,\ldots$. It is
clear that the boundary conditions given by Eq.(\ref{condition1})
correspond, in the non-relativistic limit, to the familiar
condition of a vanishing wave function at the walls of the box:
$\phi_1(0)=\phi_1(L)=0$ and the probability ($\rho$) and current
($j$) densities defined as
\begin{equation}
\rho= \overline{\phi}_1\phi_1+\overline{\chi}_2\chi_2,
\end{equation}
\begin{equation}
j=e\psi^\dagger \alpha_x \psi = ec\left(\overline{\phi}_1\chi_2
+\overline{\chi}_2^-\phi_1 \right).
\end{equation}
satisfy the following boundary conditions:
\begin{equation}
\rho(0)= \rho(L) \label{density1},
\end{equation}
\begin{equation}
j(0)= j(L)=0 \label{current1}.
\end{equation}
This implies that the particle is confined inside the box.

\section{Kicked Dirac particle confined in a one dimensional box}

Consider the relativistic spin-half particle confined in a box and interacting
with the external delta-kicking potential of the form
$$
V(x,t) =-\varepsilon\cos\left(\frac{2\pi}{\lambda} x \right)\sum_l\delta(t-lT),
$$
where $\varepsilon$ and $T$ are the kicking strength and period, respectively.
The dynamics of the system is governed by the time-dependent Dirac equation
which is given as
\begin{equation}
i\frac{\partial}{\partial t} \Psi(x,t) =\left[-i\alpha_x\frac{d}{dx} +\beta (1+
V(x,t))\right]\Psi(x,t), \label{perturbed1}
\end{equation}
for which the box boundary conditions given by Eq.(\ref{condition1}) are
imposed. To avoid complications in the treatment caused by Klein paradox and
pair creation, we choose the kicking potential  as a Lorentz-scalar, i.e. as
the mass term.  Then the exact solution of Eq.(\ref{perturbed1}) can be
obtained within a single kicking period as in the Refs. \cite{Casati,Well}.
Expanding the wave function, $\Psi(x,t)$ in terms of the complete set of the
eigenfunctions given by Eq.(\ref{ps1}) as
$$
\Psi(x,t) = \sum A_n(t) \psi_n(x), \label{1111}
$$
and inserting  this expansion into  Eq.(\ref{perturbed1}), by integrating the
obtained equation within one kicking period we have
\begin{equation}
A_n(t+T)=\sum_l A_l(t) V_{ln}e^{-iE_l T}, \label{evol}
\end{equation}
where
$$
V_{ln} =\int \psi^\dagger_n(x) e^{i\varepsilon \cos (2\pi
x/\lambda)}\psi_l (x)dx,
$$
and $E_l$ are defined by Eq.(\ref{eigenvalue}). Using the relation
\begin{equation}
e^{i\varepsilon \cos x} =\sum_{m=-\infty}^{\infty} b_m(\varepsilon)e^{im x},
\end{equation}
where $b_m(\varepsilon) =i^m J_m(\varepsilon)$, the matrix elements
$V_{ln}$ can be calculated exactly and analytically. It is clear that the norm
conservation in terms of expansion coefficients, $A_n(t)$ reads as
$$
N(t) = \sum_n |A_n(t)| =1,
$$
that follows from
$$
\int_{0}^{L}|\Psi(x,t)|^2 dx =1\;\; {\rm and}\;
\int_{0}^{L}\psi_m^\dagger(x)\psi_n(x)dx =\delta_{mn}.
$$
In choosing of the initial conditions, i.e. the values of $A_n(0)$ one should
use the norm conservation.

Having found $A_n(t)$,  we can calculate any dynamical characteristics of the
system such as average kinetic energy, probability density, wave packet
evolution, etc. The average kinetic energy as a function of time can be written
as
$$
\langle E(t)\rangle =  \int \Psi^\dagger(x,t)\left(-i\alpha_x \cdot
\frac{d}{dx}\right)\Psi(x,t)dx ,
$$
where $A_n(t)$ are given by Eq.(\ref{evol}).

\begin{figure}[htb]
\includegraphics[trim=0.5cm 0cm 0.8cm 0cm, clip=true, width=42mm]{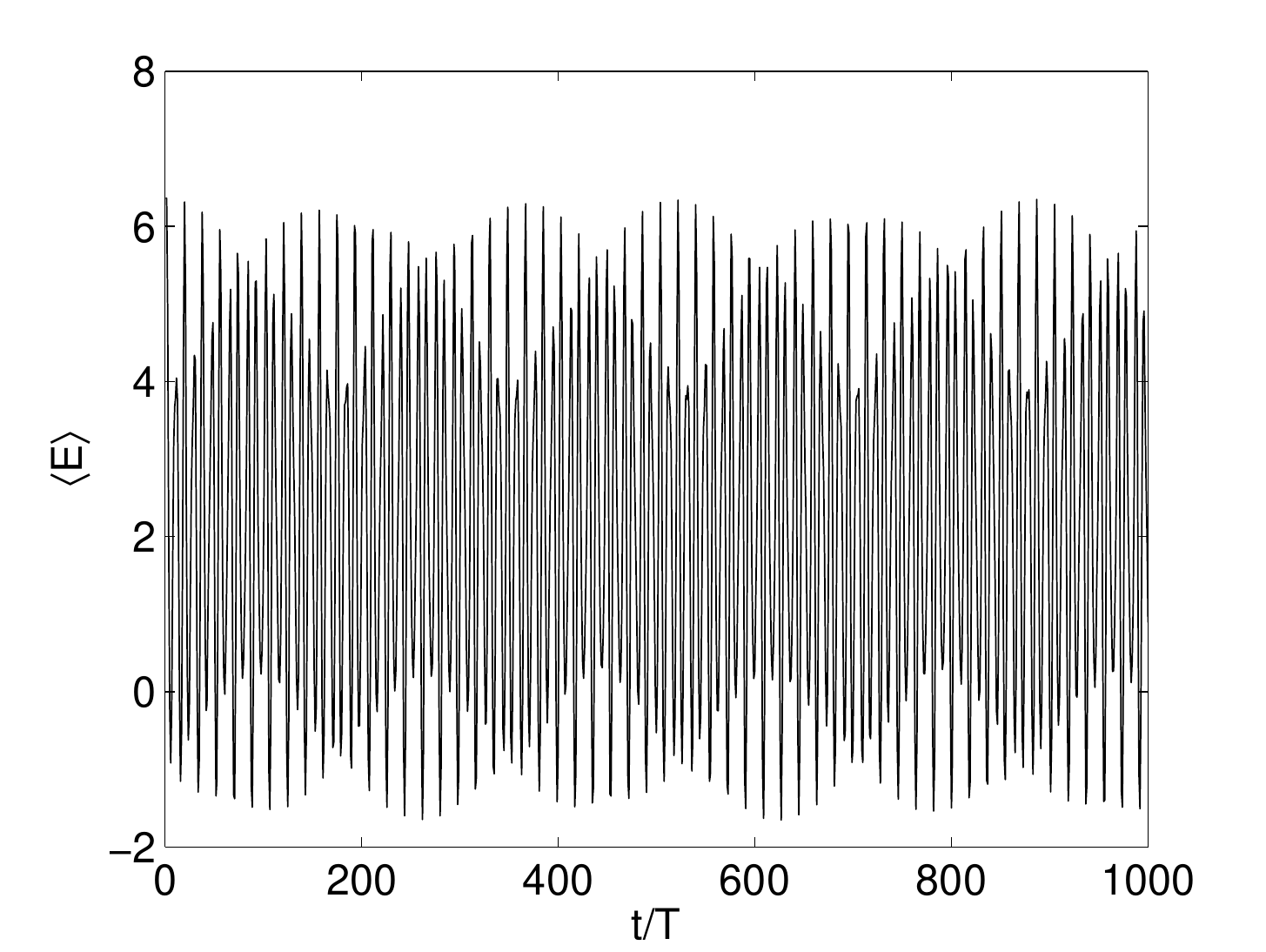}
\includegraphics[trim=0.4cm 0cm 0.8cm 0cm, clip=true, width=42mm]{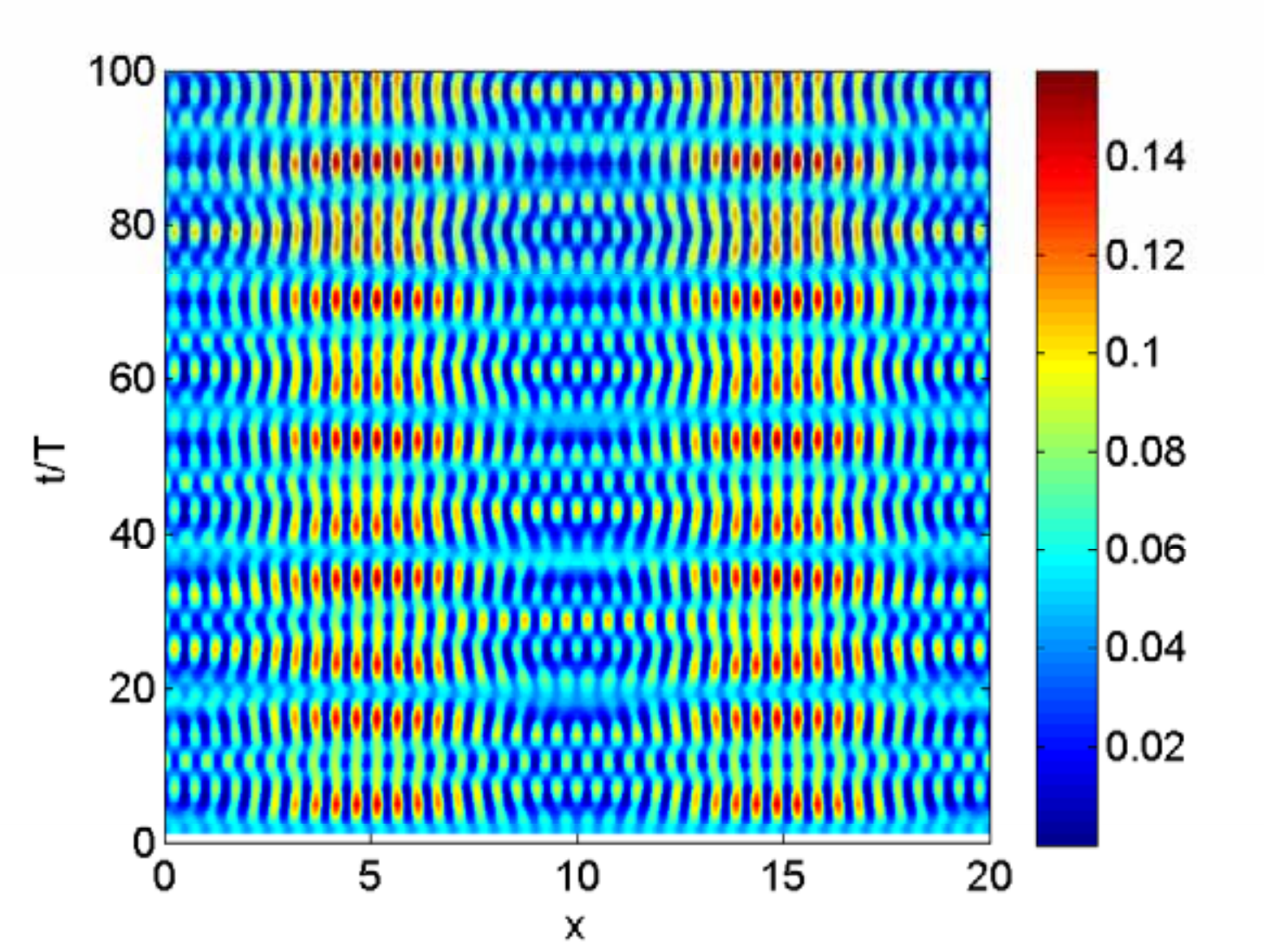}
\caption{(Color online) Time-dependence of the average kinetic
energy (a) for the value of the kicking strength $\varepsilon =
0.5$ and period $T=200$  and the corresponding probability density
plot (b)for the time interval $[0\ 100T]$. The initial state
chosen as the 40th positive energy level of the unperturbed
system.} \label{energ1}
\end{figure}

\begin{figure}[htb]
\includegraphics[trim=0.5cm 0cm 0.8cm 0cm, clip=true, width=42mm]{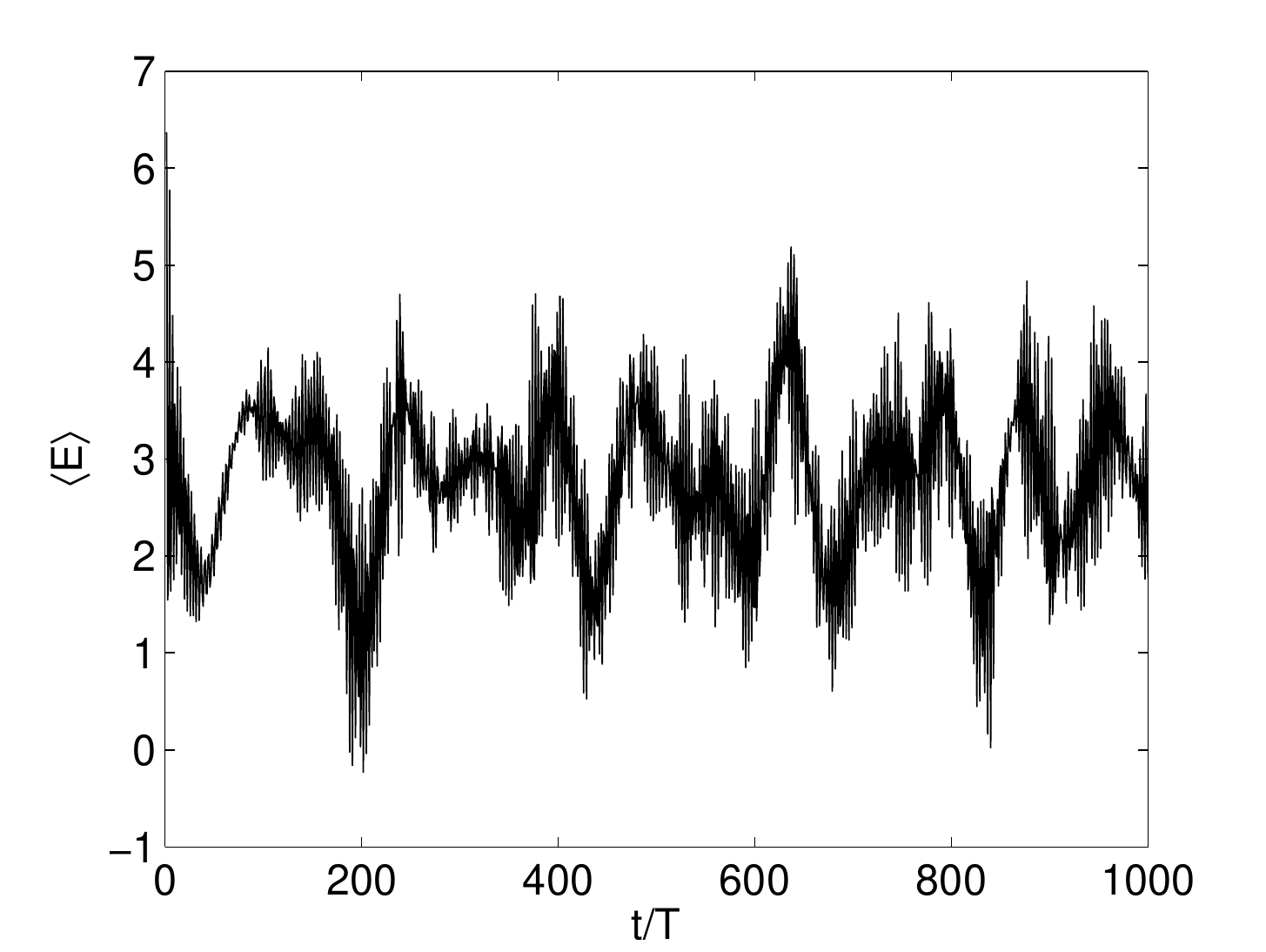}
\includegraphics[trim=0.5cm 0cm 0.8cm 0cm, clip=true, width=42mm]{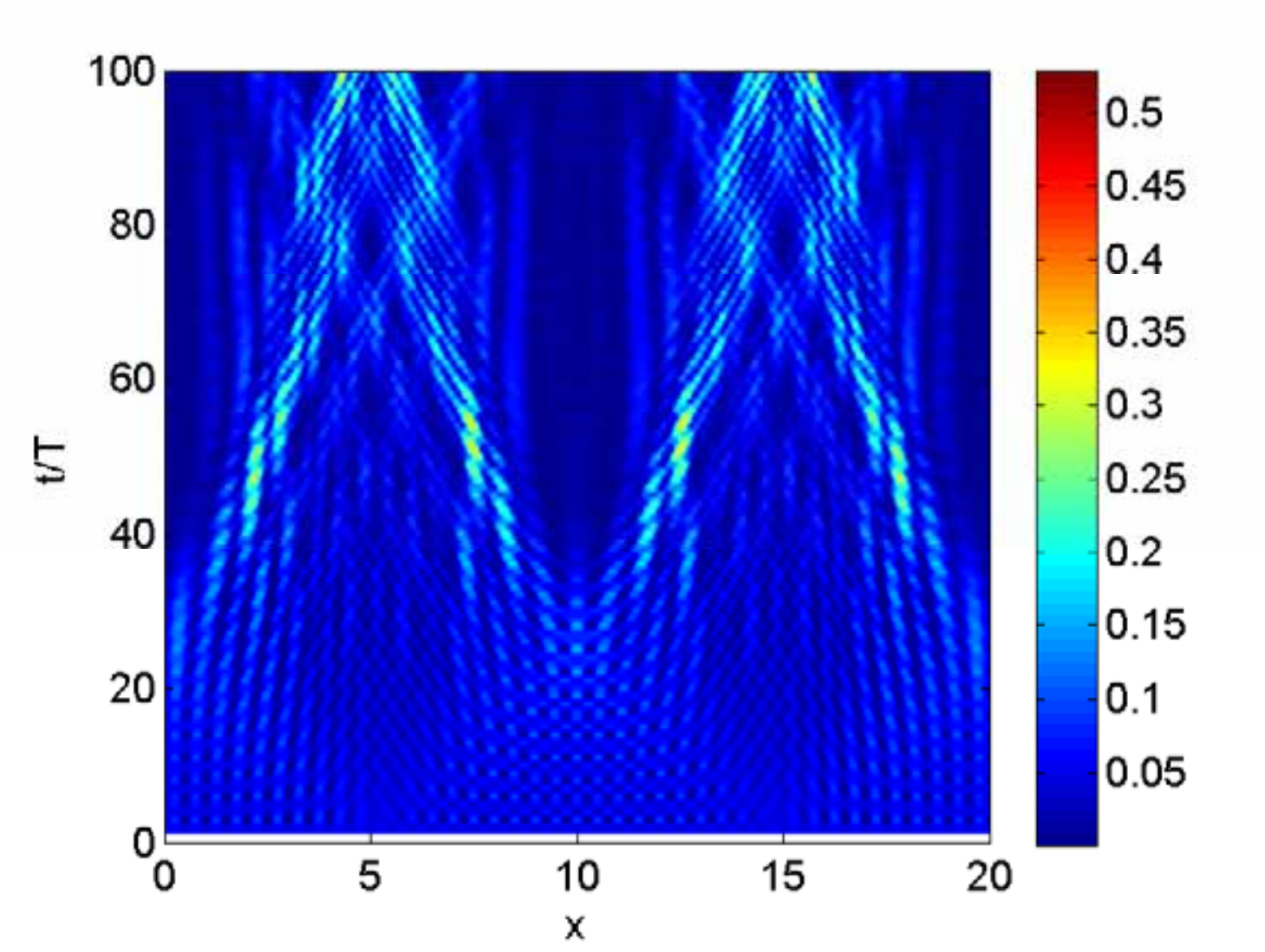}
\caption{(Color online) Time-dependence of the average kinetic
energy (a) for the value of the kicking strength $\varepsilon = 1$
and period $T=10$  and the corresponding probability density plot
(b) for the time interval $[0\ 100T]$. The initial state chosen as
the 40th positive energy level of the unperturbed system.}
\label{energ2}
\end{figure}

For the non relativistic counterpart of our model, non
relativistic kicked particle in a box, the average classical
kinetic energy grows linearly in time, while in quantum case such
a growth is suppressed \cite{Well}. This is caused by so-called
quantum localization effect \cite{Casati,Izr}. The latter implies
that no unbounded acceleration of a nonrelativistic kicked quantum
particle is possible (except the special cases of quantum
resonances \cite{Casati,Izr}). Here we explore behavior of
$\langle E(t)\rangle$ in the relativistic case for the system
described by Eq.(\ref{perturbed1}) for different kicking regimes.
In Fig.\ref{energ1} time-dependence of the average kinetic energy
(a) and spatio-temporal evolution of the probability density (b)
are plotted for $\varepsilon=0.5$ and $T=200$. Time-dependence of
the average kinetic energy shows the quasi-periodic behavior. Such
behavior is confirmed by the plot of probability density which is
also quasi-periodic both in space and time. Fig.\ref{energ2}
presents time-dependence of the average kinetic energy (a) and
probability density (b) at the values of the kicking parameters
$\varepsilon = 1$ and $T=10$. The time dependence of $E(t)$ is
neither periodic, nor monotonic and fluctuates around some value
in this kicking regime. However, the plot of the probability
density shows that particle is mostly localized in two areas
within the box. Depending on the position of the particle in the
box, the kicking potential can be attractive or repulsive, which
caused by the presence of the factor $\cos x$. Therefore, the
behavior of $E(t)$ in this system depends on the localization
particle's position inside the box and the synchronization of
kicking and bouncing (at the wall) regimes. If the particle's
motion is localized in the area where the kicking potential is
positive, the particle gains energy. Localization of the motion in
the area, where the kicking potential is negative causes the loss
of energy by particle. If the particle's motion is
time-periodically localized in these two areas, $E(t)$ will be
time periodic (or quasiperiodic). For the regime when such
localization time is different for the areas where the kicking
potential is attractive and positive, the periodicity of $E(t)$ is
broken and we have time dependence of the average kinetic energy
presented in Fig.\ref{energ2}.

\section{Zitterbwegung and wave packet dynamics}

In this section we consider dynamics of the kicked Dirac particle in a box by
exploring its transport properties. Very interesting feature of the Dirac
particle is so called trembling motion (zitterbwegung), which is the pure
relativistic effect. It was first studied by Schr\"{o}dinger
\cite{Schrod,Barut,Bjor,Thaller}, who showed that  the free relativistic
electron experiences trembling motion in vacuum. Later it was shown that such
motion can occur in the interaction of the electron with external field.
Trembling motion is characterized by the time-dependence of the average
position which is given by

$$
\langle x(t)\rangle =\langle \Psi(x,t)|x|\Psi(x,t) \rangle.
$$

Here we calculate the average position  by choosing the initial wave function
in the form of spinor Gaussian wave packet which can be written as \cite{Max1}:
\begin{equation}
\Psi(x,0)=\frac{f(x)}{\sqrt{|s_1|^2+|s_2|^2+|s_3|^2+|s_4|^2}}\left\{
\begin{array}{c}  s_1 \\ s_2 \\ s_3 \\ s_4  \end {array} \right\},
\label{initialwp}
\end{equation}
where $s_1,s_2,s_3$ and $s_4$ determine the initial spin polarization and
$$
f(x)=\frac{1}{d\sqrt{\pi}}\exp\left[ - \frac{(x-x_0)^2}{2d^2}+iv_0x\right].
$$

Then the initial conditions in terms of the wave packet can be written as:
\begin{equation}
A_n(0)=\int_0^L \psi^\dagger_n(x)\Psi(x,0) dx .
\end{equation}

\begin{figure}[htb]
\includegraphics[width=80mm]{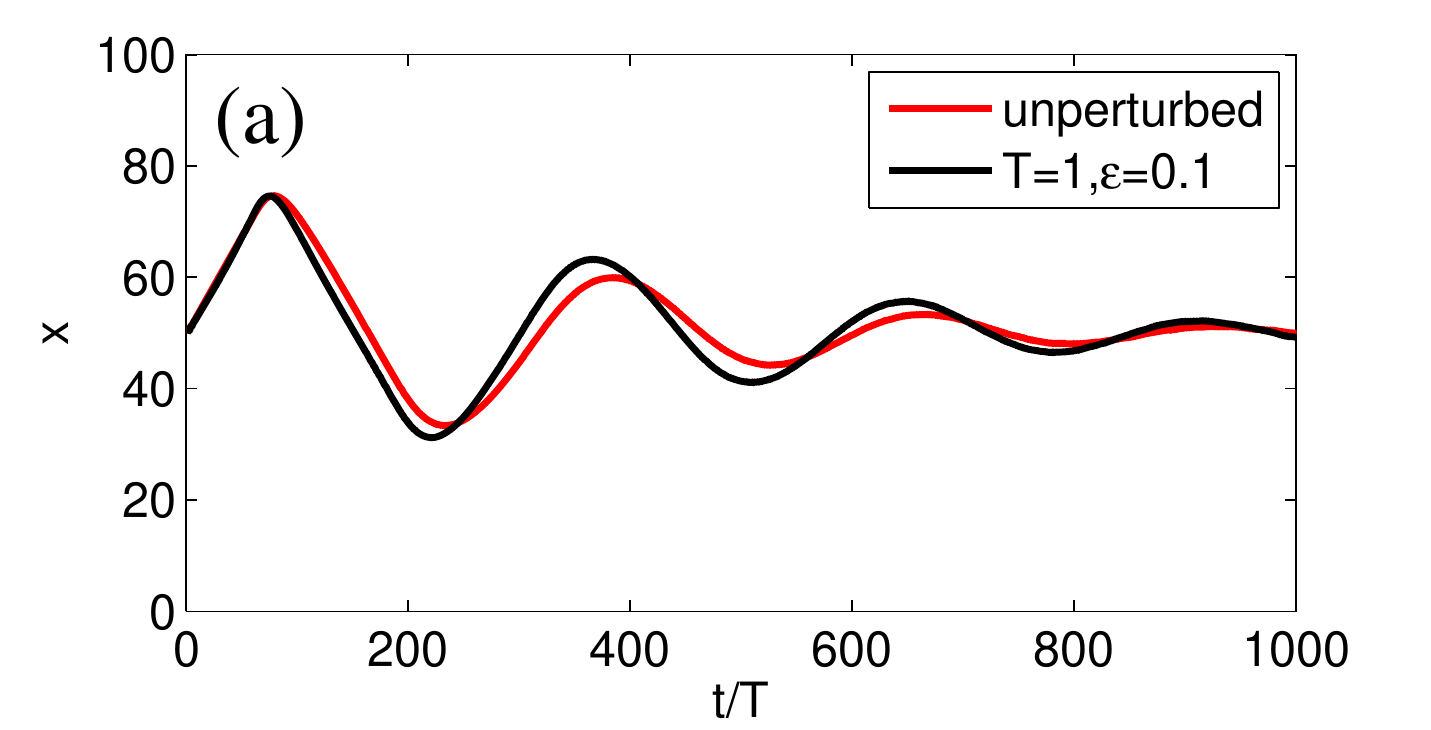}
\includegraphics[width=80mm]{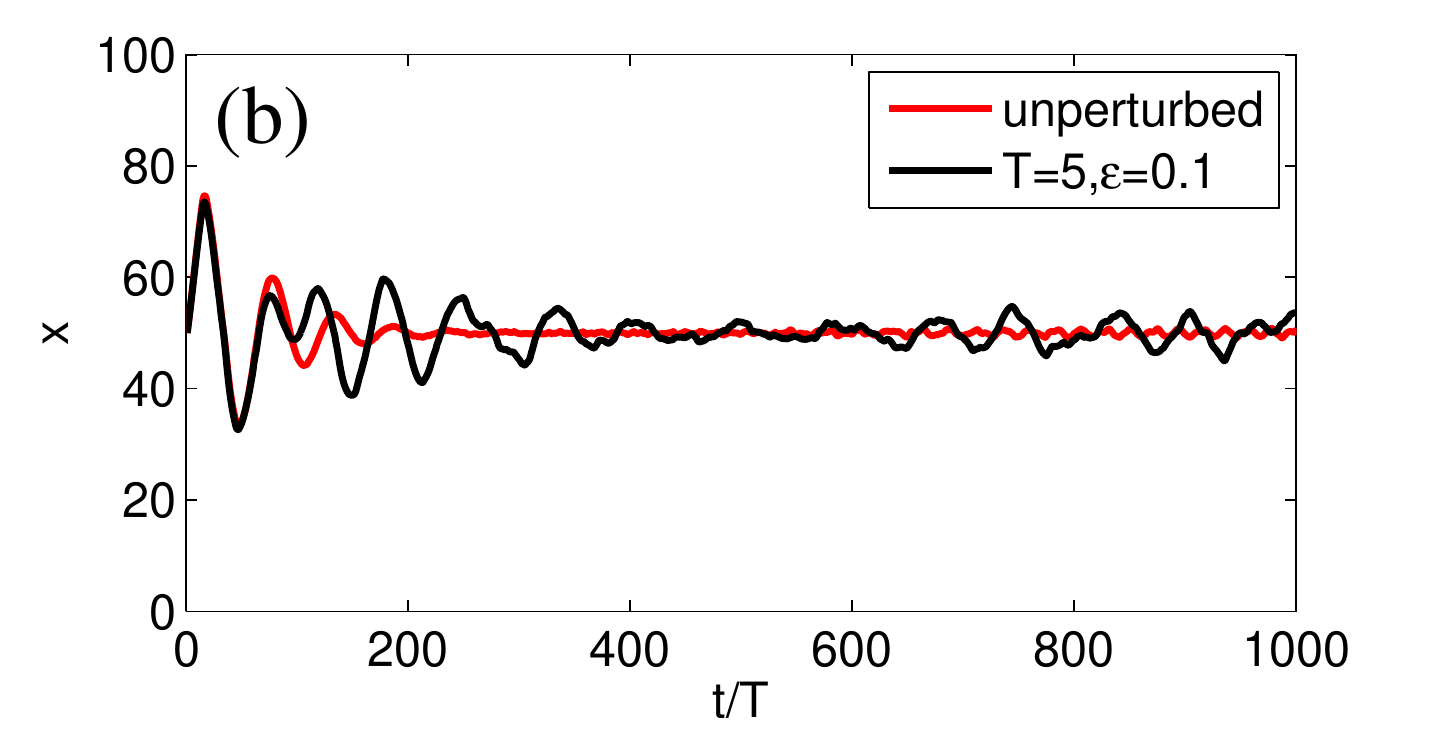}
\includegraphics[width=80mm]{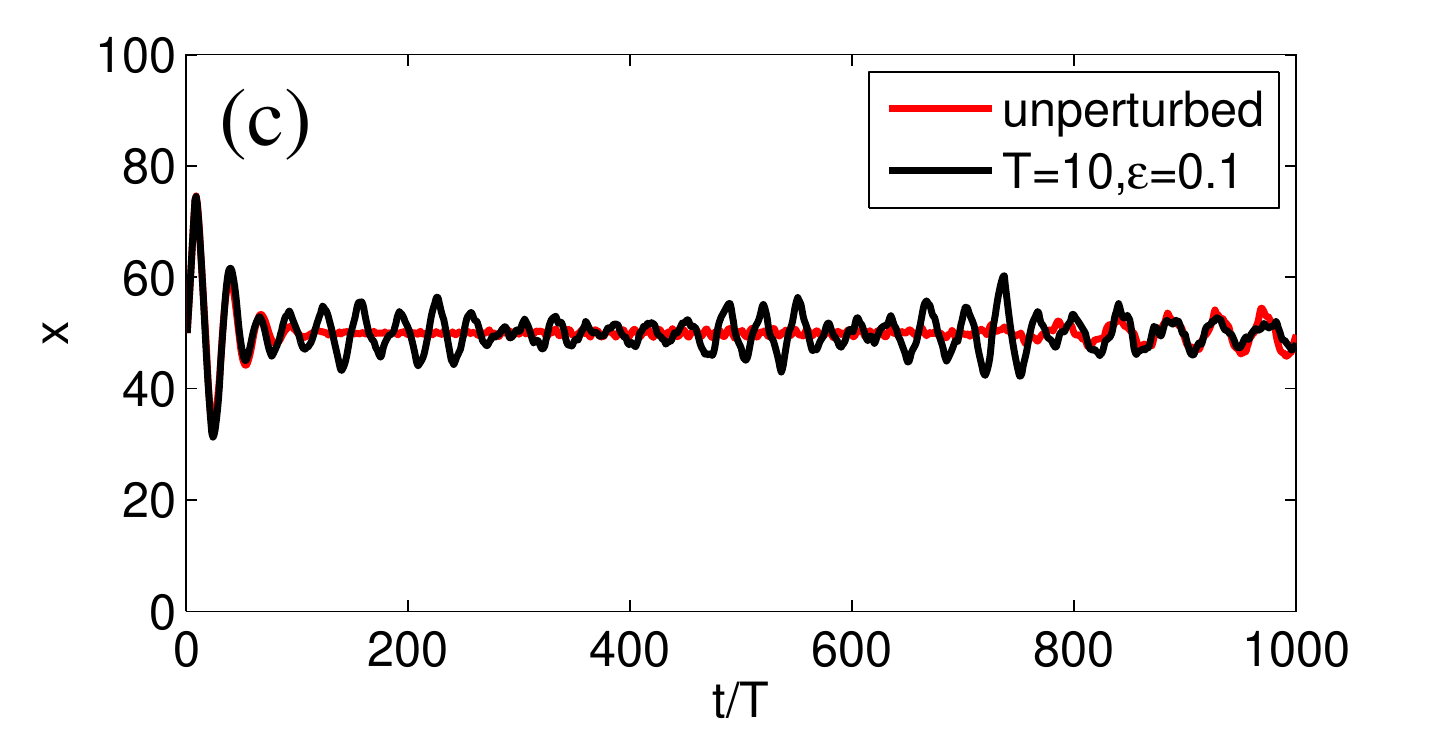}
\caption{(Color online) Trembling motion for unperturbed and kicked systems for fixed value of the  kicking strength $\varepsilon=0.1$ at different values of the kicking period: (a) $T=1$; (b)
$T=5$; (c) $T=10$. The Gaussian wave packet is initially localized at $x=L/2$,
initial group velocity is $v_0=-0.75$} \label{zitt}
\end{figure}

\begin{figure}[htb]
\includegraphics[width=80mm]{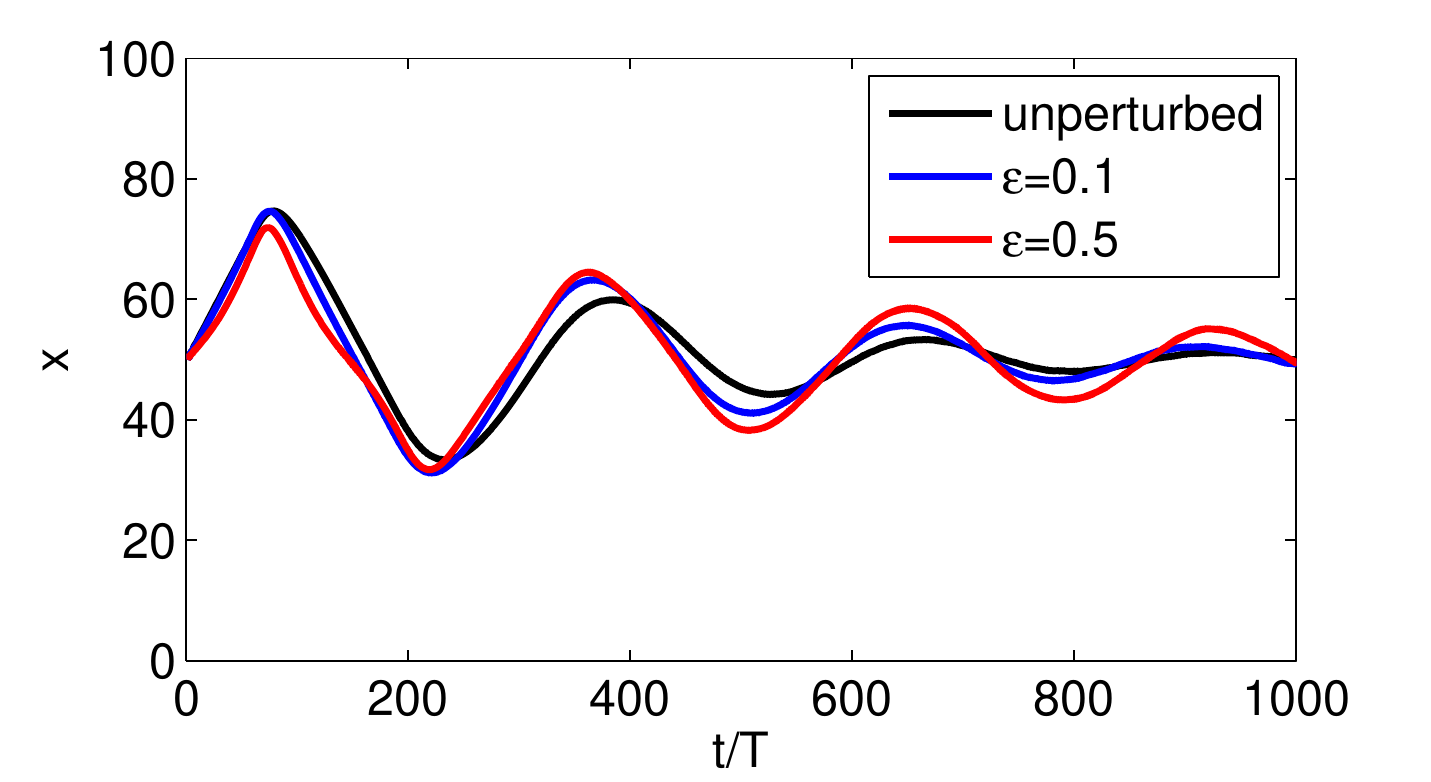}
\caption{(Color online) Trembling motion for unperturbed and kicked systems for different values of the kicking strength at fixed $T=1$. The Gaussian wave
packet is initially localized at $x=L/2$, initial group velocity is
$v_0=-0.75$} \label{zitt1}
\end{figure}

In Fig.\ref{zitt} the average position of the kicked Dirac
particle in a box is plotted for fixed $\varepsilon = 0.1$ at
different values of the kicking period, $T$ and compared with that
of unperturbed system. It can be seen from these plots that for
unperturbed particle damping of motion occurs much faster compared
to kicked one. Fig.\ref{zitt1} presents similar plots for fixed
$T$ at different values of the kicking parameter, $\varepsilon =
0.5$. The profiles of the average position are close to each
other. Therefore we may conclude that the trembling dynamics is
more sensitive to the change of $T$ rather than $\varepsilon$.
Such sensitivity can be explained by existence (unlike, e.g.,
kicked rotor) of two factors acting on the trembling motion,
bouncing and kicks. Synchronization or de-synchronization of these
factors play important role in the dynamics of the particle. Thus
in case of the unperturbed system trembling becomes damped and
suppressed after some time, while for the kicked system such
suppression is not possible. In calculation of the average
position the initial wave packet parameters have been chosen as
those in the second example in the section IV of the Ref.
\cite{thaller1}.

\begin{figure}[h]
\includegraphics[trim=2cm 0cm 1.7cm 0cm, clip=true, width=86mm]{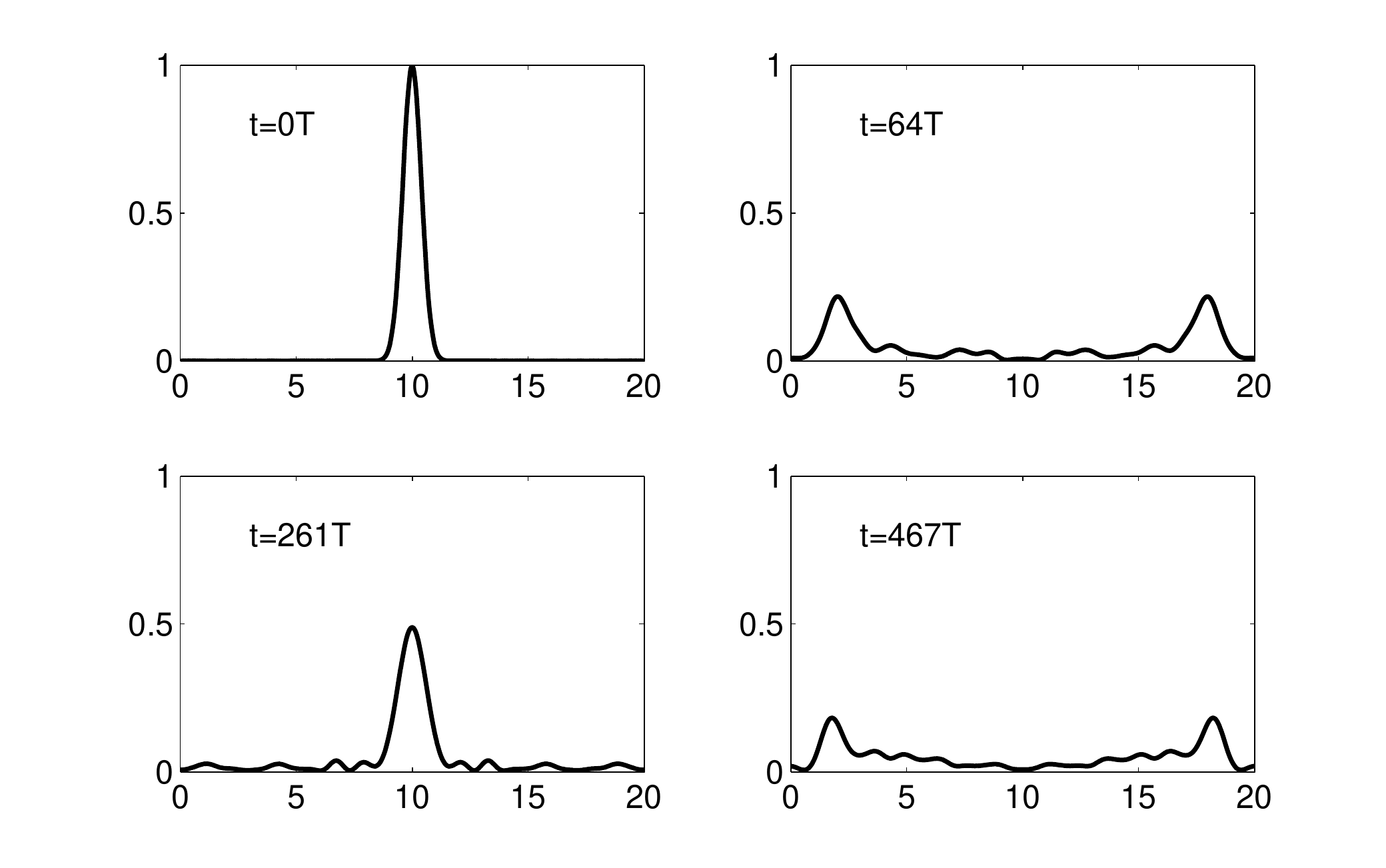}
\caption{Gaussian wave packet dynamics for $\varepsilon=0.1$ and $T=10$ at
fixed times $0$, $64T$, $261T$ and $467T$. The Gaussian wave packet with the
width $d=L/50$ is initially located at the center of the box. The initial group
velocity is $v_0=0$. $s_1=1, s_2=s_3=s_4=0$} \label{wpd}
\end{figure}

An important information about the particle transport in quantum regime can be
extracted from the wave packet evolution. Fig.\ref{wpd} presents plots of the
profile of the Gaussian wave packet (for $d=L/100 $, $x_0=L/2$ and $v_0=0$) at
different time moments ($t=0,$\;$64T,$\;$261T$ and $467T$). Dispersion of the
packet and splitting into two symmetric parts can be observed for this regime
of motion. Such splitting is caused by the existence of the spin of particle
which can have two values (up and down). Numerical experiments for different
kicking regimes showed that revival of the Gaussian wave packet is not possible
in this system.

\section{Conclusions}

We have studied classical and quantum dynamics of relativistic particle
confined in a 1D box and driven by external delta-kicking potential by
considering spin-half Dirac particle.

Classical dynamics of the system is studied by analyzing of the
discrete map describing the phase space evolution and average
kinetic energy. In particular, it is shown that for regular
dynamics the average kinetic energy does not monotonically grow by
showing fluctuations around some value, while in fully chaotic
regime the energy monotonically grows. In mixed regime, when the
phase space portrait contain both regular and chaotic
trajectories, the average kinetic energy grows during certain
number of kicks, after that one can observe suppression of the
growth.

In quantum case, the driving potential in the Dirac equation is
taken to be the Lorentz-scalar, i.e. included into the mass term.
The quantum dynamics of the system is studied by computing of the
average kinetic energy as a function of time and the
spatio-temporal evolution of the probability density. It is found
that the average kinetic energy can be quasi-periodic in time or
can fluctuate around some value. Such behavior can be explained by
the dependence of the particle dynamics on bouncing on the wall
and interaction with the kicking potential. The latter can be
attractive or repulsive depending on the position of particle
inside the wall and may cause its acceleration or deceleration.
When particle is trapped in the area where the kicking force is
attractive it looses its energy, while being trapped in the
repulsive kicking area it gains the energy. If the trapping times
for these two areas are equal, the average kinetic energy is
time-periodic. By tuning of the kicking parameters one can achieve
that the amplitude and period of $E(t)$ can be as higher and
longer as we want, i.e. the average kinetic energy can grow in
time during very long time. However, no monotonic growth of the
average kinetic energy is possible as in the case of quantum
resonances \cite{Izr} appearing in non-relativistic counterpart of
the system. The analysis of the Gaussian wave packet evolution
shows that dispersion of the packet can occur via splitting into
two symmetric part and no revival is possible. The trembling
motion of the particle is also studied by considering unperturbed
and kicked particles in a box. It is shown that damping in the
trembling motion is more sensitive to kicking period than to
kicking strength.

The above model can be used to describe Dirac particles in
graphene quantum dot, graphene nanoribbon rings and  carbon
nanotubes driven by external time-periodic fields. Kicking
potential can be realized by embedding these systems in a standing
wave in generated by optical cavities. Particle physics
realization of the kicked Dirac particle in a box comes from the
MIT Bag models of quarks where quarks confined in a bag subjected
to the influence of external time-periodic fields.

\acknowledgements{This work is partially supported by a grant of the Committee
for Coordination of Science and Technology of Uzbekistan (Ref. Nr F-2-003)}

\end{document}